\magnification=\magstep1
\baselineskip=14 pt
\hsize=5 in
\vsize=7.3 in
\pageno=1
\centerline{\bf   THERMAL AND DYNAMICAL PARTICLE CREATION} 
\centerline{\bf IN A CURVED GEOMETRY} 
\vskip 1,8cm 
\centerline{\bf Carlos E. Laciana} 
\centerline {\sl Instituto de Astronom\'{\i}a y F\'{\i}sica del Espacio}
\centerline{\sl Casilla de Correo 67 - Sucursal 28, 1428 Buenos Aires, 
Argentina} 
\centerline{\sl E-mail: laciana@iafe.uba.ar}
\bigskip
{\bf Abstract}

 A generalization of {\it Termo Field Dynamics} to a curved geometry is 
proposed. In particular a neutral scalar field minimally coupled to gravity 
is considered as matter content in a Robertson-Walker metric. A non linear 
amplification in the particle creation is obtained, due  
to the altogether action of thermal and  
geometric effects. As  a consequence the frequencies in the system  
look like red shifted with respect to the case where the thermal creation 
is not taken into account. 
 
\vskip 2,5cm
1. {\bf Introduction}

 A realistic description of the dynamics of our universe must take into 
account all  the  possible  interactions.    But  that  is practically
impossible.  An  approach  that  we can used is to separate the universe in
system and enviroment. Then we supouse degrees of freedom with different 
hierarchy, some of them considered as quantum degrees of freedom and the
other ones constitute the classical background.  The interaction between both 
kinds of degrees of freedom can be produced through the changes in the 
spacetime  which  are  driven by  the  dynamical  Einstein  equations  and  by
spontaneous  thermal  fluctuations of the enviroment  which  acts  as a  thermal
reservoir.  In the cosmological semiclassical approach  (see  ref.  [1]) the
role of the classical background is played by the gravitational field. This 
background can also play the role of thermal  reservoir,  as  is shown in ref.
[2] and the conformal quantum thermal fluctuation of the gravitational field 
can  be  interpreted  as  the  tilde fields of 
the formalism of Thermo  Field  Dynamics  (TFD)
 [3]. However the tilde fields ``inhabit" also in a curved geometry, 
then they are also affected by the changes in the metric. So we will 
consider a matter field  in  a  thermal  bath,  due  to  the  radiation
background, and also to the interaction with 
the gravitational field.  Therefore 
we can use  {\it  Quantum  Field  Theory  in  Curved  Space Time at Finite
Temperature }, or in a more realistic way, instead of finite temperature,   
a nonequilibrium description. In order to do that it is necessary to  
introduce some kind of  interaction between the system and the thermal bath.
A simple and elegant model of that interaction is proposed in the  
TFD formalism [3]. The fundamental hypothesis is:  
{\it the increase of the energy in the 
system can be produced by the excitation of an additional quanta or by 
annihilation of holes of particles  from  the  reservoir}. This is  
spontaneous process, as is stressed in ref. [4]. The other mechanism that  
increase the energy of the system is due to the  creation of particles from the 
interaction between  the  scalar  and  gravitational  fields (see ref.  [1]).
This interaction is introduced indirectly by means of the curved spacetime 
(CST). The evolution of the particle creation due to the last process is 
driven by the field equation. When the in-out situation is considered the 
Bogoliubov transformation that connect both Cauchy surfaces produces  
squeezed states as is shown in references [5] and [6]. In that sense the 
isotropic universe, taking this work into account, is another example, 
as the ones given in ref. [2], that relate the dynamics to the squeezed 
states. 
        
 The TFD hypothesis, as is shown in ref. [3] is implemented 
introducing  an  extended  Fock  space  that includes the
tilde  states  $|\tilde n>$ of the reservoir  with  the  states  $|n>$  of  the
system. Then we have the extended Fock space given by: 
$$\{|n,{\tilde n}>\}=\{|n>\}\otimes\{|{\tilde n}>\}$$       
\vskip 0,2 cm
In particular we use for the vacuum the notation $|{\bf 0}>=|0,{\tilde 0}>=
|0>|{\tilde 0}>$, with the operators $a_{\bf k}$, ${{\tilde  a}_{\bf  k}}$, 
${{a_{\bf k}}^{\dagger}}$ and ${{{\tilde a}^{\dagger}}_{\bf k}}$, such that 
$$a_{\bf k}|{\bf 0}>=0,\ \ \ {{\tilde  a}_{\bf  k}}|{\bf 0}>=0$$        
$${{a_{\bf k}}^{\dagger}}|{\bf 0}>=|1_{\bf k}>,\ \  etc\ \ \ \ \ 
{{{\tilde a}^{\dagger}}_{\bf k}}|{\bf 0}>=|{\tilde 1}_{\bf k}>,\ \ etc.$$
\vskip 0,2 cm
Moreover the operators satisfy the commutation relations 
$$[{a_{\bf k}},{{a_{\bf k^{\prime}}}^{\dagger}}]=
{{\delta}_{{\bf k},{\bf k^{\prime}}}},\ \ \    
[{{\tilde a}_{\bf k}},{{{\tilde a}^{\dagger}}_{\bf k^{\prime}}}]=
{{\delta}_{{\bf k},{\bf k^{\prime}}}}$$
\vskip 0,2 cm

 In the formulation given in ref.[7] the tilde field is a fictitious one, 
it does not represent  any  physical  magnitude. 
In  ref.  [8]  Israel  found  a
parallelism  between this approach and the problems where an event horizont 
is present, like  for  example  the  Rindler  observer  and  the  black  hole
radiation. Then the role of tilde fields is played by physical but hidden 
modes.  In ref.  [2] the conformal fluctuationof the metric 
was proposed as tilde field .  The  real  role 
of the tilde field surely must be played by a
more  complicated  combination  of  the  different   fields  present  in  the
background. The use of one particular candidate can be interpreted as 
the choice a 
kind of coarse graining in the model of the system in a thermal bath.
In the present work we can obtain interesting conclusions without 
specifying the nature of the tilde fields.  
 In order to calculate statistical mean values of the physical observables, 
in TFD  (see  ref. [3]),  the thermal  vacuum state  is  introduced.  If
$T=T({\beta})$ is the temperature of the bath, the thermal state is defined by 
the transformation         
$$|{\bf 0}, \beta>=U[{a_{\bf k}},\ {{\tilde  a}_{\bf  k}},\   
{{a_{\bf k}}^{\dagger}},\ {{{{\tilde a}_{\bf k}}^{\dagger}}}] |{\bf 0}>\eqno
(1.1)$$
\vskip 0,2 cm
 
 The operator $U$ allows us to obtain the thermal 
vacuum as a mixture of states at 
zero temperature. Then the fundamental hypothesis of TFD can be imposed by 
the expression:
$$a|{\bf  0}, \beta>= {\lambda}{{{\tilde  a}^{\dagger}}}|{\bf  0},\beta>\eqno
(1.2)$$
\vskip 0,2 cm
 
 In  general  for  a system  out  of  equilibrium,  the  operators  are  time
dependent and related with the latest ones by transformations as follows  
(see ref. [9]): 
$$a(t)={S^{-1}}(t)aS(t)$$
$${a^{\dagger\!\dagger}}(t)={S^{-1}}(t){a^{\dagger}}S(t)$$ 
\vskip 0,2 cm
Where the $\dagger\!\dagger$ symbol means that in general 
${(a^{\dagger\!\dagger})^{\dagger}}\not=a$ because $S$ is not necessarily  
unitary. The eqs (1.2) turn therefore to 
$$a(t)|{\bf 0},\beta(t)>={\lambda}(t){{\tilde a}^{\dagger\!\dagger}}(t)
|{\bf 0}, \beta (t)>\eqno (1.3a)$$
$$<{\bf 0}, \beta (t)|{a^{\dagger\!\dagger}}(t)=\gamma(t)<{\bf 0}, \beta (t)|
{\tilde a}(t)\eqno (1.3b)$$
\vskip 0,2 cm
We can use in the following the notation of ref.  [9], i.e.  :  $\lambda =
F$, $\gamma=fF^{-1}$.   

 In TFD  the  statistical  mean  value  of some physical observable is the  
thermal  vacuum  expectation    value  of  the  operator  associated  to  the
observable (see ref.   [6]).    In  particular  the  mean  value  of  created
particles will be 
$${n_{t\beta}}(t)=<0,\beta|{a^{\dagger\!\dagger}}(t)a(t)|0,\beta>\eqno (1.4)$$
\vskip 0,2 cm

 The eqs (1.3) motivate the definition of the operators $a(\beta)$ and 
${a^{\dagger}}(\beta)$ such that 
$$a(\beta)|0,\beta>=0\eqno (1.5a)$$
$$<0,\beta|{a^{\dagger}}(\beta)=0\eqno (1.5b)$$
\vskip 0,2 cm
with $a(\beta (t))\propto (a(t)-F{{\tilde a}^{\dagger\!\dagger}}(t))$ and 
${a^{\dagger}}(\beta (t))\propto ({a^{\dagger\!\dagger}}(t) - f{F^{-1}})$. 
Following refs [2] and [4] we can choose $f$ and the proportionality constant 
such that:  
$${a(\beta(t))\choose {{\tilde a}^{\dagger}}(\beta(t))}=
{(1+{n_{t \beta}})^{1/2}}\pmatrix {1&{-F}\cr {-fF^{-1}}&1}
{a(t)\choose {{\tilde a}^{\dagger\!\dagger}}(t)}\eqno (1.6)$$
\smallskip 
 
 $f$,  $F$  and  ${n_{t  \beta}}$ are also time dependent functions.  In
order to calculate ${n_{t \beta}}$ by means of eq.(1.4), we need perform the 
inverse operation to transformation (1.6), i.e.: 

$${a(t)\choose {{\tilde a}^{\dagger\!\dagger}}(t)}=
{(1+{n_{t \beta}})^{-{1/2}}}{(1-f)^{-1}}\pmatrix {1&{F}\cr {fF^{-1}}&1}
{a(\beta(t))\choose {{\tilde a}^{\dagger}}(\beta(t))}\eqno (1.7)$$
\smallskip 
\vskip 0,2 cm
2.{\bf Quadrivectorial notation}
\vskip 0,1 cm 
It will be convenient to introduce a quadrivectorial notation, in order to 
include also  the  $a^{\dagger}$  and  ${\tilde  a}$  operators.  Then we can
define (see ref. [10]): 
 
$${{\bf A}_{\bf k}}:=\pmatrix{{a_{\bf k}}\cr 
                                    {{a_{\bf k}}^{\dagger}}\cr 
                                      {{\tilde a}_{\bf k}}\cr
      {{{\tilde a}^{\dagger}}_{\bf k}}\cr}\eqno (2.1)$$
\vskip 0,2 cm
 
Let us also introduce the  $4\times4$ matrix  ${\bf \Upsilon}$ so 
that the transformation can be written as 

$${{\bf A}_{\bf k}}(\beta(t))=
{{\bf \Upsilon}_{\bf k}}(\beta(t)){{\bf A}_{\bf k}}(t)\eqno (2.2a)$$
\vskip 0,2cm
$${{\bf A}_{\bf k}}(t)=
{{{\bf \Upsilon}_{\bf k}}^{-1}}(\beta(t))
{{\bf A}_{\bf k}}(\beta(t))\eqno (2.2b)$$
\vskip 0,2cm

where 
\vskip 0,1cm

$${{\bf \Upsilon}_{\bf k}}(\beta (t))={(1+{n_{t\beta}})^{1/2}}
{\pmatrix{ {\bf I}&{\bf L}
\cr {\bf L}&{\bf I}\cr}}\eqno (2.3)$$
$${\bf L}=\pmatrix{ {0}&{{-F}}\cr {{-f{F^{-1}}}}&{0}\cr}$$ 
$${{{\bf \Upsilon}_{\bf k}}^{-1}}={(1+{n_{t\beta}})^{-1/2}}{(1-f)^{-1}}
{\pmatrix{ {\bf I}&{-{\bf L}}
\cr {-{\bf L}}&{\bf I}\cr}}\eqno (2.4)$$
Moreover ${\bf I}$ is the $2\times2$ identity matrix.

 From the aplication of the transformation (1.7) in eq. (1.4), we obtain 
$$n_{t{\beta}}(1+{n_{t{\beta}}})=f(1-f)^{-2}$$
\vskip 0,2 cm
then the relation between $n_{t{\beta}}$ and $f$ is: 
$${n_{t{\beta}}}={f\over {1-f}}\eqno (2.5)$$
\vskip 0,2 cm

 It is interesting to note two things: 
\item { a)} $n_{t{\beta}}$ does not depend on $F$. 
\item { b)} The Planckian spectrum corresponds to a particular $f$ function; 
$f=exp ({-{\beta}\epsilon})$ with $\epsilon$ the energy by mode.  
\vskip 0,2 cm
3.{\bf Dynamical and thermal effects in CST}
\vskip 0,1 cm

 The change  in  time  of  the  operators  is  driven  by  the dynamical
equations. These equations, in our case,  has the form of Klein-Gordon 
ones in a curved 
space time.  The main difference with flat space time is that the coefficients 
in a normal mode expansion are time dependent. In order to obtain the time 
dependence of the annhilation-creation operators we will use the 
method developed by  Parker  [11]  about which we will present  
a brief review.  In our
case the system and reservoir is considered in a curved geometry.  The system 
is for us a massless and neutral scalar field minimally coupled to the 
gravitational field, given by a Robertson-Walker metric. 

 The action for the matter field is 
$$S={1\over 2}\int {\sqrt {-g}} d^{4}x{\partial_{\mu}}{\varphi}
{\partial^{\mu}}{\varphi} \eqno (3.1)$$
\vskip 0,2 cm
and the metric 
$$ds^{2}=dt^{2}-{a(t)^{2}}(dx^{2}+dy^{2}+dz^{2})\eqno (3.2)$$
\vskip 0,2 cm
then the variation of the action gives the field equation 
$${\bigtriangledown}_{\mu}{\partial}^{\mu}{\varphi}=0\eqno (3.3)$$
\vskip 0,2 cm
(with ${\mu}=0,1,2,3$).      

 In order  to make the calculation easier, we can use a discretization in the
same form as in [11], i.e. we can introduce the periodic boundary condition 
$\varphi ({\bf x}+{{\bf n}L},t)=\varphi ({\bf x},t)$, 
where ${\bf n}$ is a vector 
with integer Cartesian components and $L$ a length which goes to infinity 
at the end  of  the  calculation.    Then  the  integral $\int d^{3}k$ can be
replaced by $(2\pi /L^{3})\sum_{\bf  k}$.    In  order  to  expand  a general
solution of the field equation we can introduce the set of functions 
$\{{\phi_{\bf k}}(x)\}\bigcup\{{{\phi^{\ast}}_{\bf k}}(x)\}$ defined by 
$$\phi_{\bf k}(x)={1\over {{{(La(t))}^{3/2}}{\sqrt {2W}}}} 
\exp i\big({\bf k}.{\bf x}-
{\int^{t}}_{t_{0}}W(k,{t^{\prime}}){dt^{\prime}}\big),\eqno (3.4)$$
\vskip 0,2cm
where $W$, a real function of $k=|{\bf k}|$ and $t$, 
has to be determined by the 
field equations and  the boundary conditions.  Then we can expand the field 
as   

$$\varphi({\bf x},t)={\sum_{\bf k}}[a_{\bf k}(t){\phi_{\bf k}}(x) + 
{a^{\dagger}}_{\bf k}(t){{\!\phi^{\ast}}_{\bf k}}(x)]=
{\sum_{\bf k}}[a_{\bf k}{\psi_{\bf k}}(x) + 
{a^{\dagger}}_{\bf k}{{\!\psi^{\ast}}_{\bf k}}(x)]\eqno (3.5)$$
\vskip 0,2 cm
with $a_{\bf k}:=a_{\bf k}(t=t_{1})$ when $t\geq t_{1}\geq t_{0}$ and 
with (as in ref. [11])
$$\psi_{\bf k}(x)=h({\bf k},t) \exp {i {\bf k}.{\bf x}}\eqno (3.6)$$  

$$h({\bf k},t)={1\over {{{(La(t))}^{3/2}}{\sqrt {2W}}}}
[{{\alpha ({\bf k},t)}^{\ast}}e^{-i{\int^{t}}_{t_{0}}Wdt^{\prime}}+
{{\beta ({\bf k},t)}^{\ast}}e^{i{\int^{t}}_{t_{0}}Wdt^{\prime}}]\eqno (3.7)$$
\vskip 0,2 cm
$\alpha$ and  $\beta$  are  kwown  as  Bogoliubov  coefficients.  In order 
 the normalization condition, holds for the base of field equation solutions,  
those coefficients must satisfy   
  $$|{\alpha}|^{2}-|{\beta}|^{2}=1\eqno (3.8)$$
\vskip 0,2 cm
Eq. (3.8) is equivalent to the parametrization of the coefficients 
in the form: 

$$\alpha  ({\bf  k},t)=e^{-i\gamma_{\alpha}({\bf  k},t)}  \cosh  {\theta({\bf
k},t)}\eqno (3.9a)$$
$$\beta ({\bf k},t)={e^{i{\gamma_{\beta}}({\bf k},t)}} \sinh {\theta({\bf
k},t)}\eqno (3.9b)$$
\vskip 0,2 cm
Replacing eqs (3.5)-(3.9) in the field eq. (3.3) we have the system of eqs
$$(1+\cos \Gamma \tanh \theta)M+2W{{\dot \gamma}_{\alpha}}=0\eqno (3.10a)$$
$$M{\sin \Gamma}+2W{\dot \theta}=0\eqno (3.10b)$$
\vskip 0,2 cm
where we define $\Gamma={{\gamma}_{\alpha}}+{{\gamma}_{\beta}}-
2{{\int^{t}}_{t_{0}}}W{dt^{\prime}}$, and 
$$M=-{1\over 2}{({\dot W}/W\dot)}+{1\over 4}{({\dot W}/W)^{2}}-
{9\over 4}{({\dot a}/a)^{2}}-{3\over 2}{({\dot a}/a\dot)}+
{{k^2}\over {a^2}}-{W^{2}}\eqno (3.11)$$

 From eqs (3.5)-(3.11) we can obtain the Bogoliubov transformation between 
the operators given by 

$${a_{\bf  k}}(t)={e^{i{{\gamma}_{\alpha}}({\bf k}, t)}}\cosh {{\theta}({\bf
k},t)}{a_{\bf  k}}+{e^{i{{\gamma}_{\beta}}({\bf k},  t)}}\sinh {{\theta}({\bf
k},t)}{a^{\dagger}}_{-{\bf k}}$$
  
$${{a^{\dagger}}_{\bf  k}}(t)={e^{-i{{\gamma}_{\beta}}({\bf    k}, t)}}
\sinh {{\theta}({\bf
k},t)}{a_{\bf  k}}+{e^{-i{{\gamma}_{\alpha}}({\bf k}, t)}}\cosh {{\theta}({\bf
k},t)}{a^{\dagger}}_{-{\bf k}}\eqno (3.12)$$ 

 Then we can define the mean value of created particles 
between the Cauchy surfaces labeled by the times $0$ and $t$, in the form 
$$n_{t,0}(t) = 
<{\bf 0}|{{a_{\bf k}}^{\dagger}}(t)a_{\bf k}(t)|{\bf 0}>\eqno (3.13)$$

 Using the inverse transformation of eqs (3.12) it is easier to see that 
$$n_{0,t}(t)=
<{\bf 0},t|{{a_{\bf k}}^{\dagger}}a_{\bf k}|{\bf 0},t>\eqno (3.14)$$
\vskip 0,2 cm
Moreover from eqs (3.10) and (3.13) we obtain that $n$ can be 
written in the form 
$n_{0,t}=(g-1)^{-1}$ with $g$ a function of the particle model used, i.e.:
$$g={{\cos^{2}\Gamma}\over {\big(1+2{W\over M}{{\dot
\gamma}_{\alpha}}\big)^{2}}}\eqno (3.15)$$
\vskip 0,2 cm
The function $g$ plays a  role analogous to that 
of $f^{-1}$ from eq. (2.5). For 
some particle model eq. (3.15) permits us to obtain Planckian spectrum.  

 In the quadrivectorial notation the transformation given by eqs (3.12) is 
  
$${{\bf A}_{\bf k}}(t)={{\bf \Omega}_{\bf k}}(t){{\bf A}_{\bf k}}\eqno (3.16)$$
\vskip 0,2 cm
with
$${\bf \Omega}=\pmatrix{ {\bf U}&{\bf 0}\cr {\bf 0}&{\bf U}\cr}$$   
$${\bf U}=\pmatrix{  {e^{{i{\gamma}_{\alpha}}({\bf  k},t)}{\cosh{\theta}({\bf
k},t)}},
&{{e^{{i{\gamma}_{\beta}}({\bf  k},t)}{\sinh{\theta}({\bf
k},t)}}{\bf P}
}\cr {{e^{{-i{\gamma}_{\beta}}({\bf  k},t)}{\sinh{\theta}({\bf
k},t)}}{\bf P}, 
}&{{e^{{-i{\gamma}_{\alpha}}({\bf  k},t)}{\cosh{\theta}({\bf
k},t)}}
}\cr}$$

 ${\bf P}$ is a parity operator, such that 
$${\bf P} a_{\bf k} = a_{-{\bf k}}$$
\vskip 0,2 cm
and ${\bf 0}$ is the 2x2 zero matrix. 
 The transformation given by the matrix ${\bf \Omega}$ is unitary (see 
ref. [11]). Then in our case ${\dagger\!\dagger} = {\dagger}$ because the 
dissipation is not introduced. Moreover as the dynamical change of the 
operators is supposed independent of the thermal one, so we have 
$${{\bf A}(t,{\beta}(t))}={{\bf \Upsilon}({\beta (t)})}{\bf \Omega}(t){\bf A}:=
{{\bf \Lambda}(t)}{\bf A}\eqno (3.17)$$

 therefore 

$${\bf \Lambda}={(1+{n_{0\beta}})^{1/2}}{(1+{n_{t0}})^{1/2}}\pmatrix{{\bf R}& 
{\bf T}\cr
{\bf T}&{\bf R}}$$ 
\vskip 0,1cm
with 
$${\bf R}=\pmatrix{e^{i{{\gamma}_{\alpha}}}&{e^{i{{\gamma}_{\beta}}}}
{\tanh \theta}{\bf P}\cr {e^{-i{{\gamma}_{\beta}}}}{\tanh \theta}{\bf P}& 
e^{-i{{\gamma}_{\alpha}}}},$$
$${\bf T}=\pmatrix{-F{e^{-i{{\gamma}_{\beta}}}}{\tanh \theta}{\bf P}&-F
e^{-i{{\gamma}_{\alpha}}}\cr -f{F^{-1}}e^{i{\gamma}_{\alpha}}&-f{F^{-1}}
{e^{i{{\gamma}_{\beta}}}}{\tanh \theta}{\bf P}}$$ 

 By means of the last transformation we can reobtain $n_{t,{\beta}}$ in the 
form 
$${n_{t,{\beta}}}=
<{\bf 0}|{a^{\dagger}}(t,{\beta})a(t,{\beta})|{\bf 0}>\eqno (3.18)$$
\vskip 0,2 cm

 It is easy to prove that also the following equality is valid: 
$$<{\bf 0},t,{\beta}|{a^{\dagger}}a|{\bf 0},t,{\beta}>=
<{\bf 0}|{a^{\dagger}}(t,{\beta})a(t,{\beta})|{\bf 0}>\eqno (3.19)$$
\vskip 0,2 cm

 Performing the calculation of $n_{t,{\beta}}$, using the transformation 
(3.17), results (as it is shown in ref. [10]): 
$$n_{t{\beta}}=n_{t0}+n_{0{\beta}}+2n_{t0}n_{0{\beta}}\eqno (3.20)$$
\vskip 0,2 cm
As is mentioned in ref.[10], the increment in the number of created 
particles, due to the temperature is equal to the one produced in curved space 
at zero temperature, when there is a particle distribution in the initial 
state, as is shown in ref.[11]. That is due to the analogous role of the 
operators ${{\tilde a}_{\bf k}}$ to the $a_{-{\bf k}}$ in curved space (see 
ref.[12]). A similar amplification effect in the particle creation was 
obtained in ref.[13] with a massive scalar field interacting with a 
massless  scalar field in a Robertson-Walker metric. 
\vskip 0,2 cm
4.{\bf Thermal equilibrium}

 We will see now what the effect in the thermal distribution of the 
field modes is, due to the presence of the thermal reservoir. In order to do 
that the equilibrium situation is considered. Then let us extremize the 
Helmholtz free energy 
$$F=E-{1\over {\beta}}K\eqno (4.1)$$
\vskip 0,2 cm
Where $E$ is the statistical mean energy at time $t$  and temperature 
$T=T({\beta})$, given by 
$$E=<{\bf 0},t,{\beta}|{{\hat H}(t)}|{\bf 0},t,{\beta}>\eqno (4.2)$$
\vskip 0,2 cm
whith $\hat H$ the hamiltonian operator. $K$ is the entropy, which is 
obtained as 
$$K=<{\bf 0},t,{\beta}|{{\hat K}(t)}|{\bf 0},t,{\beta}>\eqno (4.3)$$  
\vskip 0,1cm
where the $\hat K$ operator was introduced in ref.[7] in the form 
$${\hat K}=-{\sum_{\bf k}}\{{{a^{\dagger}}_{\bf k}}{a_{\bf k}} log\  
{n_{t{\beta}}}({\bf k}) - {a_{\bf k}}{{a^{\dagger}}_{\bf k}} 
log\  (1+ {n_{t{\beta}}}({\bf k}))\}\eqno (4.4)$$
\vskip 0,2cm
Therefore we obtain the entropy of a Bose gas  (see ref.[14]), given by 
$$K={-\sum_{\bf k}}\{{n_{t{\beta}}}({\bf k})log\ {n_{t{\beta}}}({\bf k})-
(1+{n_{t{\beta}}}({\bf k}))log\ (1+{n_{t{\beta}}}({\bf k}))\}\eqno (4.5)$$
\vskip 0,1 cm
with $n_{t{\beta}}$ as in eq.(1.4). 

 We can calculate the energy, as in  ref.[9],  from the tensorial  
energy-momentum operator 
$${{\hat  T}_{\mu\nu}}={\sum_{{\bf  k}{\bf  k}^{\prime}}}
\{{a_{\bf  k}}{a_{{\bf k}^{\prime}}}
{D_{\mu\nu}}[{\psi_{\bf k}}(x),{\psi_{{\bf k}^{\prime}}}(x)] + 
{a_{\bf  k}}{{a^{\dagger}}_{{\bf
k}^{\prime}}}{D_{\mu\nu}}[{\psi_{\bf k}}(x),{{\psi^\ast}_{{\bf k}^
{\prime}}}(x)] +$$ 
$${{a^{\dagger}}_{\bf  k}}{a_{{\bf k}^{\prime}}}
{D_{\mu\nu}}[{{\psi^\ast}_{\bf k}}(x),{\psi_{{\bf k}^{\prime}}}(x)] + 
{{a^{\dagger}}_{\bf  k}}{{a^{\dagger}}_{{\bf k}^{\prime}}}
{D_{\mu\nu}}[{{\psi^\ast}_{\bf k}}(x),{{\psi^\ast}_{{\bf k}^{\prime}}}(x)] +
\{{\bf k}\leftrightarrow {{\bf k}^{\prime}}\}\}\eqno (4.6)$$
\vskip 0,1 cm
with
$$D_{\mu\nu}[\varphi,\psi]={1\over 2}{{\partial_{\mu}}{\varphi}}
{\partial_{\nu}}{\psi}-{1\over 4}{g_{\mu\nu}}
{\partial^{\sigma}}{\varphi}{\partial_{\sigma}}{\psi}$$
\vskip 0,1 cm
Then we can calculate the hamiltonian by 
$${\hat H}=\int a^{3}d^{3}x{{\hat T}_{00}}$$
\vskip 0,1 cm
where $\int d^{3}x=L^{3}$ in the discrete approximation. 

 Instead of eq.(4.2) we can calculate the energy, in an equivalent and more 
easier way, as 
$E=<{\bf  0},\beta|{\hat    H}[a_{\bf    k}(t),
{{a_{\bf    k}}^{\dagger}(t)}]|{\bf 0},\beta>$, with   
$${\hat H}={\sum_{{\bf k}{{\bf k}^{\prime}}}}
\{a_{\bf k}(t)a_{{\bf k}^{\prime}}(t){F_{1}}+{{a^{\dagger}}_{\bf    k}}(t)
{{a^{\dagger}}_{{\bf k}^{\prime}}}(t){{F_{1}}^{\ast}}+
a_{\bf k}(t){{a^{\dagger}}_{{\bf k}^{\prime}}}(t){F_{2}}+
{{a^{\dagger}}_{\bf    k}}(t){a_{{\bf    k}^{\prime}}}(t)
{{F_{2}}^{\ast}}\}\eqno (4.7)$$
\vskip 0,1 cm
where
$$F_{1}={1\over 2}{{\dot \phi}_{\bf k}}{{\dot \phi}_{{\bf k}^{\prime}}}
 + {1\over 2}({{k^2}\over {a^2}}){{ \phi}_{\bf k}}
{{ \phi}_{{\bf k}^{\prime}}}\eqno (4.8a)$$
$$F_{2}={1\over 2}{{\dot \phi}_{\bf k}}{{\dot 
{\!\phi^{\ast}}}_{{\bf k}^{\prime}}} + 
{1\over 2}({{k^2}\over {a^2}}){{ \phi}_{\bf k}}
{{\!\phi^{\ast}}_{{\bf k}^{\prime}}}\eqno (4.8b)$$
\vskip 0,1 cm
with $\phi=\phi[W]$ given by eq. (3.4) and $h$ the Hubble coefficient that for 
R-W universes is the relative velocity of the 
scale factor,  i.  e.,  $h={\dot a}/a$.  If we suppose that $W$ satisfies the
Cauchy data that  diagonalized  the  hamiltonian at $t$ time (see ref.[15]),
the expression (4.7) turns to 
$${\hat H}={\sum_{{\bf k}{\bf {k^{\prime}}}}}W\{{1\over 2} + 
{a_{\bf k}}^{\dagger}(t)a_{{\bf k}^{\prime}}(t)\}\eqno (4.9)$$
\vskip 0,1 cm

Taking into account eqs. (1.4) and (4.1) we have  
$$F={\sum_{\bf k}}\{W({1\over 2} + {n_{t{\beta}}})
+{1\over \beta}[{n_{t{\beta}}}({\bf k})log\ 
{n_{t{\beta}}}({\bf k})-
(1+{n_{t{\beta}}}({\bf k}))log\ (1+{n_{t{\beta}}}({\bf k}))]\}\eqno (4.10)$$
\vskip 0,1 cm
In equilibrium
$$\delta F=0\eqno (4.11)$$
\vskip 0,1 cm
Then the  equilibrium condition is  
 
$${{\partial F}\over {\partial n_{t{\beta}}}}=0\eqno (4.12)$$
\vskip 0,1 cm
That gives us the usual Planckian distribution
$${n_{t{\beta}}}={1\over {{e^{\beta W}}-1}}\eqno (4.13)$$
\vskip 0,1 cm
If we call $W_{0}$ the frequency of the normal modes 
in thermal equilibrium and curved spacetime,  
when the thermal effect, in the particle creation,  is not 
considered, we can write in equilibrium  
$n_{t0}=1/(e^{\beta W_{0}}-1)$. Moreover from 
eq.(3.20) we have 
$$n_{t\beta}=n_{t0}(1+{n_{0\beta}\over {n_{t0}}} + 2n_{0\beta})$$
\vskip 0,1 cm 
For high temperature, when the approximation $W\beta<<1$ (idem for $W_{0}$ 
) is valid, taking until the second term  in the exponential expansions, 
gives  
$$W\simeq {W_{0}\over {1+{n_{0\beta}\over n_{t0}} + 2n_{0\beta}}}
\eqno (4.14)$$   
\vskip 0,1 cm
In the denominator of eq. (4.14) all the terms are positive, therefore the 
frequency $W$ looks like red shifted with respect to $W_{0}$.
\bigskip

5.{\bf Conclusions and Comments}

 The particle creation due to the thermal bath and the one coming from the 
interaction with  the  geometry,   both
mechanisms of particle creation as  we  can  see  from  eq.    (3.20), 
act symmetricaly, however the nature of 
both processes is different. One of them is a spontaneous process due to the 
thermal fluctuations of the reservoir, while the other one is produced by the 
dynamical change of  geometry.    The symmetry is due to the fact that in the
two  situations  the  vacuum    state    changes    by  means  of  Bogoliubov
transformations, because still for the  fields  in  the  curved  geometry the
field  equation  is  Klein-Gordon  like.      The    similarity  of  the  two
transformations was analized in ref.  [16].   There  we  have  seen  that the
isotropy of the spacetime produces a mirror symmetry between the modes of the 
field and one of these symmetrical spaces can be  associated  with  the  tilde
field  of TFD.  In a more complex case, when  there  is  interaction  with
other  fields,  additional  amplification  of  the  particle production can be
present (see ref. [13]), producing a ``cascade" effect. 

 From eq.(4.14)  we  can  see that the observed frequency of the system with
reservoir, with respect to the one at zero 
temperature looks like red shifted due 
to the effect of thermal creation of particles (or modes of the field).

 The calculation is identical if the field is massive, but the interpretation 
can be  different,  because  the  thermal  effect changes the redistribution of
modes of the field by the interaction with the reservoir.  Then the particles 
obtain quanta of  energy  of  the medium, but no new particle appears.  In
the massless case the simple creation of modes can be interpreted as new 
massless particles.  In a flat spacetime it is necessary to introduce some 
 symmetry breaking mechanism, in order to produce massive particles from 
the vacuum  (see  ref.  [17]).  However in curved spacetime massive particles
are  created  without  an  explicit  breaking  of  symmetry.    This  can  be
interpreted using the analogy between the vacua in the curved spacetime and 
the ones related to  accelerated  observers in flat spacetime (it is known in
the literature [1],[18], as Rindler  observer).  As it is shown in ref.  [19]
the Rindler observer has  a Planckian spectrum in his reference system while 
in the Minkowski vacuum  particles are not observed.  It is similar to
the behaviour of the electromagnetic field with respect to different reference 
systems. Actually we have treated with a toy model of electromagnetic 
the field, represented in our case by the massless scalar field.
\vskip 1,8cm
\noindent 
{\bf ACKNOWLEDGEMENTS}
\medskip

 This work was supported partially by the Consejo Nacional de Investigaciones 
Cient\'{\i}ficas y T\'ecnicas (Argentina), 
by the Buenos Aires University and by the 
European Community DG XII. 
\vskip 1cm
{\bf REFERENCES}
\medskip
\item{$[1]$} Birrell N.D. and Davies P.C.W.; {\it Quantum Fields in Curved 
Space} (Cambridge Universy Press) (1982).
\item{$[2]$} Laciana C.E.; Physica {\bf A}, {\bf 216} (1995) 511-517.    
\item{$[3]$} Umezawa H., Matsumoto H., and Tachiki M. (1982); ``Thermo 
Field Dynamics and Condensed States" (North-Holland, Amsterdam).
\item{$[4]$} Umazawa H. and Yamanaka Y.; Phys. A, {\bf 170} (1991) 291-305.  
\item{$[5]$} Grishchuk L.P. and Sidorov Y.V., Phys. Rev. D, {\bf 42}, 
$N^{o}$ 10, (1990), 3413-3421. 
\item{$[6]$} Laciana C.E.; ``Rotation operator vs particle creation in a 
curved space time" (submitted to Phys. Lett. B).
\item{$[7]$} Takahashi Y. and Umezawa H.; Collective Phenomena {\bf 2}, 
55-80, (1975).
\item{$[8]$} Israel W. (1976). Phys. Lett. {\bf 57} A, 107.
\item{$[9]$} Hardman I., Umezawa H., and Yamanaka Y.; J. Math. Phys. 
{\bf 28} (12), (1987) 2925-2938. 
\item{$[10]$} Laciana C.E.;  ``Particle creation amplification in curved space 
due to thermal effects" (preprint hep-th/9508051, submitted to the Int. 
Journ. of Modern Phys. A). 
\item{$[11]$} Parker L.; Phys. Rev. {\bf 183}, $N_{o} 5$, 1057-1068, (1969). 
\item{$[12]$} Laciana  C.E.;    Gen.    Rel.   and Grav.  {\bf 26}, $N^{o} 4$,
363-379, (1994). 
\item{$[13]$} Audretch  J.    and  Spangehl  P.;    Phys.  Rev.  D, {\bf 35},
$N^{o}$ 8, (1987), 2365-2371. 
\item{$[14]$} Laciana C.E.; ``Thermal  conditions  for  scalar  bosons in a
curved  spacetime", Gen. Rel. and  Grav.,  (in  press),\   (preprint
hep-th/9601103).  
\item{$[15]$} Castagnino M., and  Laciana C.  (1988).  J.  Math.  Phys.  {\bf
29}, 460.   
\item{$[16]$} Laciana C.E.  (1994).  Gen.    Rel.    and  Grav..    {\bf 26},
$N^{o}$ 4, 363-379.      
\item{$[17]$}  Bernstein J., Rew.  of Modern  Phys.,  {\bf  46},  $N^{o}\ 1$,
(1974). 
\item{$[18]$} Takagi S., Prog.  of Theor.   Phys.    Supplement,  $N^{o}\ 88$,
(1986). 
\item {$[19]$} Boyer T.H., Phys. Rev. D, {\bf 21}, $N^{o}\ 8$, (1980), 
pp 2137-2148. 
\bye